\documentclass[prl,aps,twocolumn]{revtex4}
\usepackage{graphicx}
\usepackage{bm}

\begin{document}

\title{Anomalous magneto-oscillations and spin precession}

\author{S. Keppeler}
\affiliation{Abteilung Theoretische Physik, Universit\"at Ulm,
Albert-Einstein-Allee 11, D-89069 Ulm, Germany}

\author{R.~Winkler}
\affiliation{Institut f\"ur Technische Physik III, Universit\"at
Erlangen-N\"urnberg, Staudtstr. 7, D-91058 Erlangen, Germany}

\date{October~17, 2001}
\begin{abstract}
  A semiclassical analysis based on concepts developed in quantum
  chaos reveals that anomalous magneto-oscillations in quasi
  two-dimensional systems with spin-orbit interaction reflect the
  non-adiabatic spin precession of a classical spin vector along the
  cyclotron orbits.
\end{abstract}


\maketitle

If in a solid the spatial inversion symmetry is broken, spin-orbit
(SO) interaction gives rise to a finite spin splitting of the energy
bands even at magnetic field $B=0$. In quasi two-dimensional (2D)
systems this $B=0$ spin splitting is frequently analyzed by
measuring the magnetoresistance oscillations at small magnetic
fields $B>0$, known as Shubnikov-de Haas (SdH) oscillations.
Following a semiclassical argument due to Onsager \cite{Ons52} it
has long been assumed \cite{EisStoNarGosWie84,gru00}, that the
frequencies $f^{\mathrm{SdH}}_\pm$ of these oscillations are
proportional to the unequal occupations $N_\pm$ of the spin-split
subbands,
\begin{equation}
\label{eq:onsager}
N_\pm = (e/2\pi\hbar) \, f^{\mathrm{SdH}}_\pm ,
\end{equation}
where $e$ is the electron charge and $\hbar$ is Planck's constant.
Recently, experiments and numerical quantum mechanical calculations
have shown that, in general, these oscillations are not simply
related to the $B=0$ spin-subband densities \cite{WinPapDePSha00}.
However, it has remained unclear when and why Onsager's
semiclassical argument fails. Here we use a semiclassical trace
formula for particles with spin, which was only lately developed
\cite{BolKep98,BolKep99a} in the context of quantum chaos, in order
to show that the anomalous magneto-oscillations reflect the
nonadiabatic spin precession along the cyclotron orbits. Currently
great efforts are made to obtain a deeper understanding of
spin-related phenomena in semiconductor quantum structures, in
particular due to possible applications in spintronics \cite{pasps}.
While spin is a purely quantum mechanical property with no immediate
analogue in classical physics, the present analysis reveals that our
understanding of spin phenomena can be greatly improved by
investigating equations of motion for a classical spin vector.

In the presence of a magnetic field perpendicular to the plane of a
2D electron system the electrons condense in highly degenerate
Landau levels that are regularly spaced in energy. With increasing
field $B$ these Landau levels are pushed through the Fermi surface
causing magneto-oscillations which reflect the oscillating density
of states (DOS) at the Fermi energy $E_F$, see,
e.g.,~\cite{AshMer76}. Onsager's semiclassical analysis of
magneto-oscillations was based on a Bohr-Sommerfeld quantization of
cyclotron orbits. However, for systems with spin there is no
straightforward generalization of Bohr-Sommerfeld quantization
\cite{LitFly91b,EmmWei96}. The Gutzwiller trace formula \cite{Gut90}
provides an alternative and particularly transparent semiclassical
interpretation of magneto-oscillations that is applicable even in
the presence of SO interaction.  Rather than giving individual
quantum energies, the trace formula relates the DOS of the quantum
mechanical system to a sum over all periodic orbits of the
corresponding classical system. As a function of energy $E$ the
individual terms oscillate proportional to $\cos[S(E)/\hbar]$ where
$S(E)$ is the action of the orbit.

First we briefly discuss magneto-oscillations of electrons with
effective mass $m^\ast$ in a 2D system without SO interaction. Here,
the sum over $k$-fold repetitions of the classical periodic
cyclotron orbits corresponds to a Fourier decomposition of the DOS
as a function of the energy $E$, where the action of the $k$-fold
revolution corresponds to the $k$th harmonic $2\pi kE /
(\hbar\omega) = 2\pi k m^\ast E/(\hbar eB)$ of the DOS with
cyclotron frequency $\omega= eB/m^\ast$. Thus we see that the DOS
for a fixed energy $E=E_F$ oscillates as a function of the
reciprocal magnetic field $1/B$ which is the origin of
magneto-oscillations. In particular, we get Onsager's formula from
the first harmonic $k=1$. Longer orbits $k>1$, giving rise to higher
harmonics in the oscillating DOS, are exponentially damped for small
but nonzero temperatures \cite{Ric00} so that here it suffices to
consider $k=1$. In general, the Gutzwiller trace formula is an
asymptotic relation that holds in the semiclassical limit $\hbar\to
0$. However, in the particular case discussed above it is an
identity.

\begin{figure}
\includegraphics[width=0.9\columnwidth]{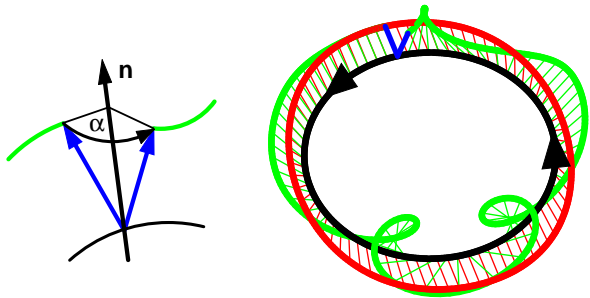}
\caption{(color) Classical spin precession (bold green line) about the
effective field $\bm{\mathcal{B}}$ (bold red line) along a cyclotron
orbit (black) for a GaAs QW. Thin lines represent the momentary
vectors of the effective field $\bm{\mathcal{B}}$ (red) and the spin
$\bm{s}$ (green) along the cyclotron orbit. The momentary vectors
for $\bm{\mathcal{B}}$ are normalized with respect to the maximum of
$|\bm{\mathcal{B}}|$ along the orbit. In the starting point we have
chosen $\bm{s} \parallel \bm{\mathcal{B}}$. After one cycle the
motion of the spin vector can be identified with a rotation by an
angle $\alpha$ about an axis $\bm{n}$, as shown in the blow-up on
the left. Initial and final directions of the spin vector $\bm{s}$
are marked in blue. The system is a 100~{\AA} wide
GaAs-Al$_{0.5}$Ga$_{0.5}$As QW grown in the crystallographic
direction [113] with 2D density $N=5\times 10^{11}$~cm$^{-2}$ in the
presence of an electric field $E_\perp =100$~kV/cm and a magnetic
field $B=0.05$~T.}
\label{spin-precession}
\end{figure}

Recently it has been shown \cite{BolKep98,BolKep99a} that in leading
semiclassical order the SO interaction results in weight factors
$2\cos (k\alpha/2)$ for the orbits in the trace formula where the
angle $\alpha$ characterizes the spin precession $\dot{\bm{s}} =
\bm{s}\times\bm{\mathcal{B}}$ of a classical spin vector $\bm{s}$
along the classical orbit \cite{alpha}. Here $\dot{\bm{s}}$ is the
time derivative of $\bm{s}$ and $\bm{\mathcal{B}}$ is an effective
magnetic field including the contributions of both SO coupling and
the Zeeman interaction due to the external magnetic field $B$ felt
by the spin along the orbit. After $k$ periods of the cyclotron
motion the spin vector $\bm{s}$ has been rotated by the angle
$k\alpha$ about an axis $\bm{n}$, see Fig.~\ref{spin-precession}. We
remark that the axis $\bm{n}$ (but not $\alpha$) depends on the
starting point of the cyclotron orbit. Like the effective field
$\bm{\mathcal{B}}$, the angle $\alpha$ depends on the external field
$B$. It contains a dynamical as well as a geometric phase similar to
Berry's phase \cite{Ber84}.  Apart from higher harmonics the
oscillating part of the DOS at the Fermi energy $E_F$ is
proportional to
\begin{equation}
\label{magneto-osc}
  \cos(\alpha/2) \: \cos\left[2\pi m^\ast E_F/(\hbar eB) \right] \, .
\end{equation}

We have analyzed magneto-oscillations for quasi 2D electron systems
in semiconductors such as GaAs where we have two contributions to
the SO coupling. The Dresselhaus term \cite{dre55a} reflects the
bulk inversion asymmetry of the zinc blende structure of GaAs. If
the inversion symmetry of the confining potential of the quasi 2D
system is broken, we get an additional SO coupling given by the
Rashba term \cite{byc84}. While the Dresselhaus term is fixed, the
Rashba SO coupling can be tuned by applying an electric field
$E_\perp$ perpendicular to the plane of the quasi 2D system
\cite{gru00}.

\begin{figure}
\includegraphics[width=1.0\columnwidth]{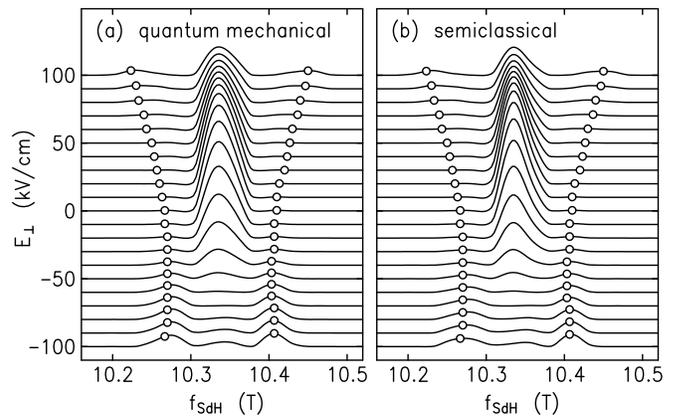}
\caption{(a) Quantum mechanical and (b) semiclassical Fourier spectra
for different values of the electric field $E_\perp$ for a 2D
electron system in a 100~{\AA} wide GaAs-Al$_{0.5}$Ga$_{0.5}$As QW
grown in the crystallographic direction [113] with constant total
density $N=5\times 10^{11}$~cm$^{-2}$. The open circles show the
expected Fourier transform peak positions $(2\pi\hbar/e)N_\pm$
according to the calculated spin subband densities $N_\pm$ at
$B=0$.}
\label{SdH-spectrum}
\end{figure}

In Fig.~\ref{SdH-spectrum} we compare the Fourier spectra of the
magneto-oscillations of the DOS calculated by means of a
diagonalization of the quantum mechanical Hamiltonian with the
spectra obtained from Eq.~(\ref{magneto-osc}) based on an
integration of the classical equations of motion for the precessing
spin. We consider here a 2D electron system in a 100~{\AA} wide
GaAs-Al$_{0.5}$Ga$_{0.5}$As quantum well (QW) grown in the
crystallographic direction [113] with constant total density $N =
N_+ + N_- = 5\times 10^{11}$~cm$^{-2}$ and with varying $E_\perp$.
For comparison, the circles mark the peak positions which one would
expect according to Eq.~(\ref{eq:onsager}) for the spin subband
densities $N_\pm$ calculated quantum mechanically at $B=0$. The
Fourier spectra are in strikingly good agreement.  On the other
hand, the peak positions deviate substantially from the positions
expected according to the $B=0$ spin splitting. In particular, the
semiclassical analysis based on Eq.~(\ref{magneto-osc}) reproduces
the central peak that is not predicted by Eq.~(\ref{eq:onsager}).
The asymmetry in Fig.~\ref{SdH-spectrum} with respect to positive
and negative values of $E_\perp$ reflects the low-symmetry growth
direction [113] (Ref.~\cite{WinPapDePSha00}).

An analysis of the classical spin precession along the cyclotron
orbit reveals the origin of anomalous magneto-oscillations. The
spin-split states at $B=0$ correspond to fixing the direction of
spin parallel and antiparallel to the effective field
$\bm{\mathcal{B}} (\bm{p})$ along the cyclotron orbit, where
$\bm{p}$ is the kinetic momentum. However, in general the precessing
spin cannot adiabatically follow the momentary field
$\bm{\mathcal{B}} (\bm{p})$. This can be seen in
Fig.~\ref{spin-precession} where we have plotted the momentary field
$\bm{\mathcal{B}}(\bm{p})$ as well as the precessing spin $\bm{s}$
along a cyclotron orbit. Both the direction and the magnitude of
$\bm{\mathcal{B}}$ change along the orbit. In particular, the
Dresselhaus term reverses the direction of $\bm{\mathcal{B}}$ when
$|\bm{\mathcal{B}}|$ has a minimum. A spin vector that is no longer
parallel or antiparallel to $\bm{\mathcal{B}}$ implies that the
system is in a superposition of states from both spin subbands so
that the magneto-oscillations are not directly related to the $B=0$
spin splitting.

For the spin, in order to be able to follow the momentary field
$\bm{\mathcal{B}} (\bm{p})$ adiabatically, the orbital motion must
be slow compared to the motion of the precessing spin, i.e., we must
have $B \ll |\bm{\mathcal{B}} (\bm{p})|$ for all points $\bf p$
along the cyclotron orbit. Therefore, it is the smallest value
$\mathcal{B}_{\mathrm{min}} = \min |\bm{\mathcal{B}} (\bm{p})|$
along the cyclotron orbit which determines whether or not the spin
evolves adiabatically. This is illustrated in
Fig.~\ref{spin-precession} where the parameters were chosen such
that initially the spin is parallel to the effective field
$\bm{\mathcal{B}}$. First $\bm{s}$ can follow $\bm{\mathcal{B}}$,
but after a quarter period of the cyclotron orbit the effective
field $\mathcal{B}$ reaches its minimum $\mathcal{B}_{\mathrm{min}}$
and $\bm{s}$ starts to ``escape'' from $\bm{\mathcal{B}}$.
Subsequently, the spin vector $\bm{s}$ is no longer parallel to
$\bm{\mathcal{B}}$ also in those regions where $\mathcal{B}$ becomes
large again. We remark that adiabatic spin precession does not imply
$\alpha=0$ but only that the rotation axis $\bm{n}$ is approximately
parallel to the initial (and final) direction of the effective field
$\bm{\mathcal{B}}$.

For many years, anomalous magneto-oscillations have been explained
by means of magnetic breakdown \cite{CohFal61,deARocBas94}.
Underlying this approach is a rather different semiclassical picture
where each spin-split subband is associated with an energy surface
with separate classical dynamics. In our treatment, on the other
hand, there is only one energy surface complemented by the dynamics
of a classical spin vector. It is the essential idea within the
concept of magnetic breakdown that in a sufficiently strong external
magnetic field $B$ electrons can tunnel from a cyclotron orbit on
the energy surface of one band to an orbit on the energy surface of
a neighboring band separated from the first one by a small energy
gap. For spin-split bands the separation of these bands is
proportional to the effective field $\mathcal{B}$, i.e., magnetic
breakdown occurs most likely in regions of a small effective field
$\mathcal{B}$. This approach implies that the anomalous
magneto-oscillations are essentially determined by the breakdown
regions only. (These breakdown regions can be identified with mode
conversion points \cite{LitFly91b}.) We want to emphasize that here
our approach differs fundamentally from these earlier models: In the
present ansatz spin continuously precesses along the cyclotron
orbit, i.e., the angle $\alpha$ in Eq.\ (\ref{magneto-osc}) is
affected by the nonadiabatic motion of $\bm{s}$ in the regions of
both small and large $\mathcal{B}$ (see Fig.~\ref{spin-precession}).

In the adiabatic regime $B \ll \mathcal{B}_{\mathrm{min}}$ the angle
$\alpha$ is given by the integral of the modulus of the momentary
field $\bm{\mathcal{B}}$ along the orbit plus a Berry phase
\cite{Ber84},
\begin{equation}
\label{adiabtic}
  \alpha = \int_0^T |\bm{\mathcal{B}}| \, \mathrm{d} t + \alpha_B \, , 
\end{equation}
where $T=2\pi/\omega$ is the period of the cyclotron motion. In the
limit of small external fields $(B\to 0)$ the Berry phase $\alpha_B$
converges towards a constant and the integrand in Eq.\ 
(\ref{adiabtic}) can be expanded with respect to a small Zeeman
term, $|\bm{\mathcal{B}}| \approx \mathcal{B}_0 + \mathcal{B}_1 B$,
where the coefficients $\mathcal{B}_0$ and $\mathcal{B}_1$ are
$T$-periodic in time. Thus in the limit of small external fields we
obtain $\alpha (B) \approx \alpha_0 / B + \alpha_1$ with constants
$\alpha_0$ and $\alpha_1$ independent of $B$. Inserting the last
relation in Eq.~(\ref{magneto-osc}) we thus retrieve
Eq.~(\ref{eq:onsager}), i.e., only in the limit of adiabatic spin
precession magneto-oscillations are directly related to the $B=0$
spin splitting. By changing the crystallographic growth direction of
the QW, it is possible to tune the value of
$\mathcal{B}_{\mathrm{min}}$. In particular, for a QW grown in the
crystallographic direction $[110]$ the Dresselhaus term vanishes for
$\bm{p}$ parallel to the in-plane directions $[001]$ and
$[00\overline{1}]$.  Thus for a symmetric QW without Rashba SO
coupling we have $|\bm{\mathcal{B}} (\bm{p})|=B$ for these values of
$\bm{p}$ which implies that there is no adiabatic regime and one
always observes anomalous magneto-oscillations.

For a system with Rashba SO coupling but no Dresselhaus term both
the classicsal equations of motion for the precessing spin and the
quantum mechanical problem can be solved analytically. Here the
Gutzwiller trace formula exactly reproduces the quantum mechanical
density of states for $B>0$. For this system, the effective magnetic
field $\bm{\mathcal{B}}(\bm{p})$ along the cyclotron orbit has a
constant magnitude $\mathcal{B}$ and for small external fields $B\to
0$ the exact solution turns into the adiabatic solution so that we
get no anomalous magneto-oscillations in agreement with an earlier
quantum mechanical analysis \cite{WinPapDePSha00}.

Finally, we note that the concepts developed here are rather general
and, in particular, are not restricted to spin-1/2 systems. Indeed,
an analogous semiclassical analysis can be carried out for any
system with (nearly) degenerate subbands. These bands can be
identified with a single band with a SO coupling acting on an
effective spin degree of freedom similar to Lipari and Baldareschi's
treatment \cite{lip70} of the multiply degenerate valence band edge
in semiconductors with diamond or zinc blende structure. In
particular, we expect that our approach can be applied to the
interpretation of de Haas-van Alphen experiments on ultrahigh-purity
samples \cite{edd82} that had called in question the established
concepts of magnetic breakdown.

SK was supported by the Deutsche Forschungsgemeinschaft (DFG) under
contract no.\ Ste 241/10-1.


\begin{thebibliography}{20}
\bibitem[{alp()}]{alpha}
\bibinfo{note}{The relation between the angle $\alpha$ and the previously
  \cite{BolKep99a} defined angles $\theta$ and $\eta$ is given by
  $\cos(\alpha/2) = \cos(\theta/2)\cos\eta$.}
\end{thebibliography}

\end{document}